\documentclass[10pt, conference, letterpaper]{IEEEtran}
\IEEEoverridecommandlockouts
\usepackage{cite}
\usepackage{amsmath,amssymb,amsfonts}
\usepackage{algorithmic}
\usepackage{graphicx}
\usepackage{textcomp}
\usepackage{xcolor}
\usepackage{kotex}
\def\BibTeX{{\rm B\kern-.05em{\sc i\kern-.025em b}\kern-.08em
    T\kern-.1667em\lower.7ex\hbox{E}\kern-.125emX}}
\begin{document}

\title{Sequential Semantic Generative Communication for Progressive Text-to-Image Generation}

\author{\IEEEauthorblockN{1\textsuperscript{st} Hyelin Nam}
\IEEEauthorblockA{\textit{School of EEE} \\
\textit{Yonsei University}\\
Seoul, Korea \\
hlnam@ramo.yonsei.ac.kr}
\and
\IEEEauthorblockN{2\textsuperscript{nd} Jihong Park}
\IEEEauthorblockA{\textit{School of Information Technology} \\
\textit{Deakin University}\\
VIC 3220, Australia \\
jihong.park@deakin.edu.au}
\and
\IEEEauthorblockN{3\textsuperscript{rd} Jinho Choi}
\IEEEauthorblockA{\textit{School of Information Technology} \\
\textit{Deakin University}\\
VIC 3220, Australia \\
jinho.choi@deakin.edu.au}
\and
\IEEEauthorblockN{4\textsuperscript{th} Seong-Lyun Kim}
\IEEEauthorblockA{\textit{School of EEE} \\
\textit{Yonsei University}\\
Seoul, Korea \\
slkim@ramo.yonsei.ac.kr}
}

\maketitle

\begin{abstract}
This paper proposes new framework of communication system leveraging promising generation capabilities of multi-modal generative models. Regarding nowadays' smart applications, successful communication can be made by conveying the perceptual meaning, which we set as text prompt. Text serves as a suitable semantic representation of image data as it has evolved to instruct an image or generate image through mutli-modal techniques, by being interpreted in a manner similar to human cogitation. Utilizing text can also reduce the overload compared to transmitting the intact data itself.
The transmitter converts objective image to text through multi-model generation process and the receiver reconstructs the image using reverse process. Each word in the text sentence has each syntactic role, responsible for particular piece of information the text contains. For further efficiency in communication load, the transmitter sequentially sends words in priority of carrying the most information until reaches successful communication. Therefore, our primary focus is on the promising design of a communication system based on image-to-text transformation and the proposed schemes for sequentially transmitting word tokens. Our work is expected to pave a new road of utilizing state-of-the-art generative models to real communication systems.
\end{abstract}


\begin{figure}[t]
    \centering
    \includegraphics[clip,width=\columnwidth]{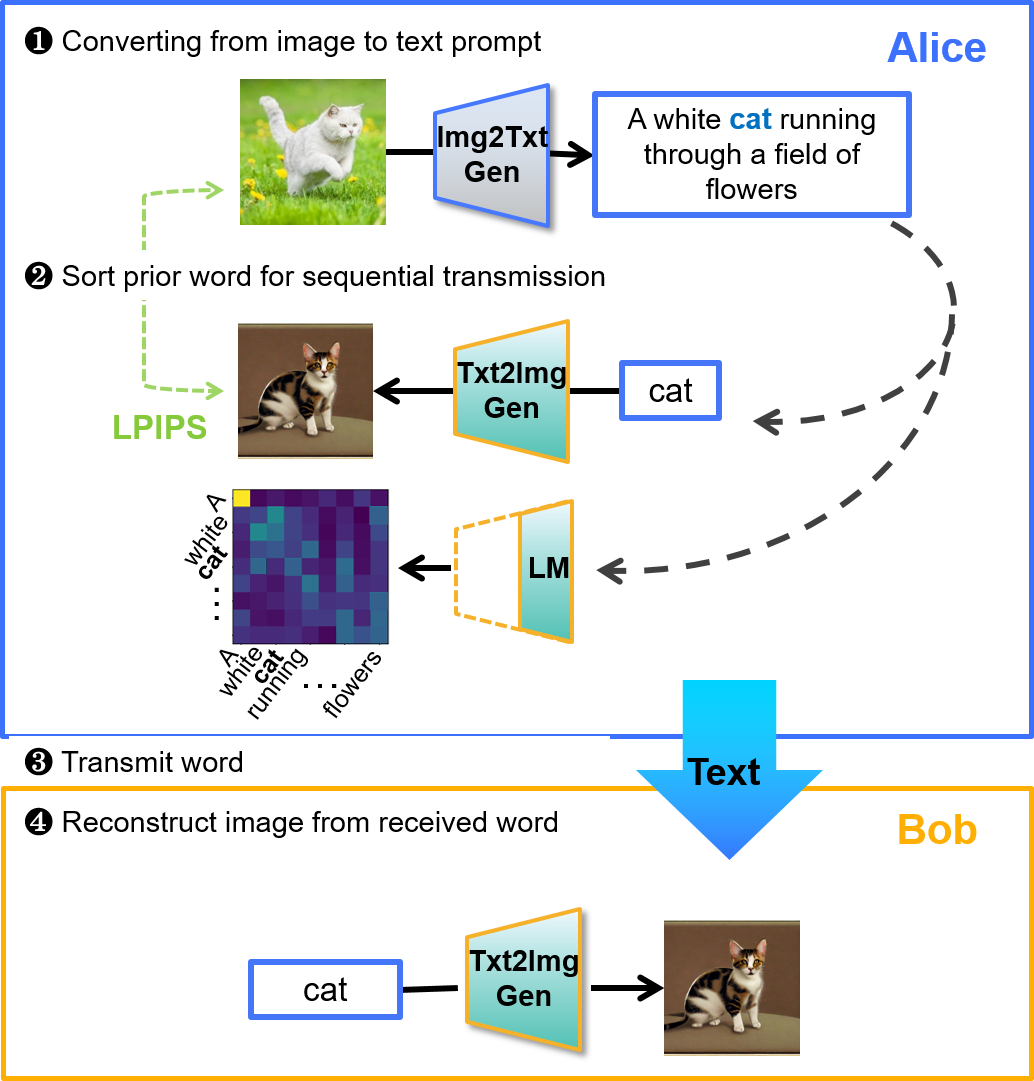}
    \caption{A schematic illustration of semantic sequential communication.}
    \label{fig: systemmodel}
\end{figure}

\section{Introduction}
Artificial intelligence (AI) plays a significant role in the recent advancements of communication systems, which face challenges such as limited bandwidth, varying channels, and diverse user requirements. In the 3GPP communication standard, discussions are actively underway to apply various deep learning techniques to communication systems, aiming to go beyond 5G and prepare for the next-generation 6G communication, with the expectation of intelligent, reliable, and high-performance systems. 

With such conspicuous advance, more diverse applications or systems have appeared in various fields of industry such as autonomous vehicles, Mobile Edge Computing (MEC) or robotics. These recent technologies require fast exchange of data and additive process to convert to usable information. This raises an importance of communicating with essential information brings more efficiency in communication system.

Recently, generative models (DALL-E\cite{ramesh2021zero}, CLIP\cite{radford2021learning}, BLIP\cite{li2022blip}) have made significant progress in multi-modal generation tasks. They understand the common semantics between two different domains of data, text and image. This paper aims to observe the expected synergistic effects between generative models and effective communication systems.  

Communication systems utilize impressive ability to interpret and explain different modalities, and generate an interactive perceptual `meaning'. Text prompt is a proper type of semantic representation to transmit in that it can compactly contain information with much smaller load than image pixels, and can be interpreted by individual word units. 

In this regard, we propose a new communication framework, where intelligent communication users successfully convey semantic meaning in the form of textual representations, and convert them to images. We fully understand promising multi-modal generative models and come up with various ways to apply to designing effective communication frameworks. As a result, we can sort transmitting data in the order of amount of important information it contains, and achieve in transmitting meaning with at least communication load.

\begin{figure*}[t]
    \centering
    \includegraphics[clip,width=0.9\textwidth]{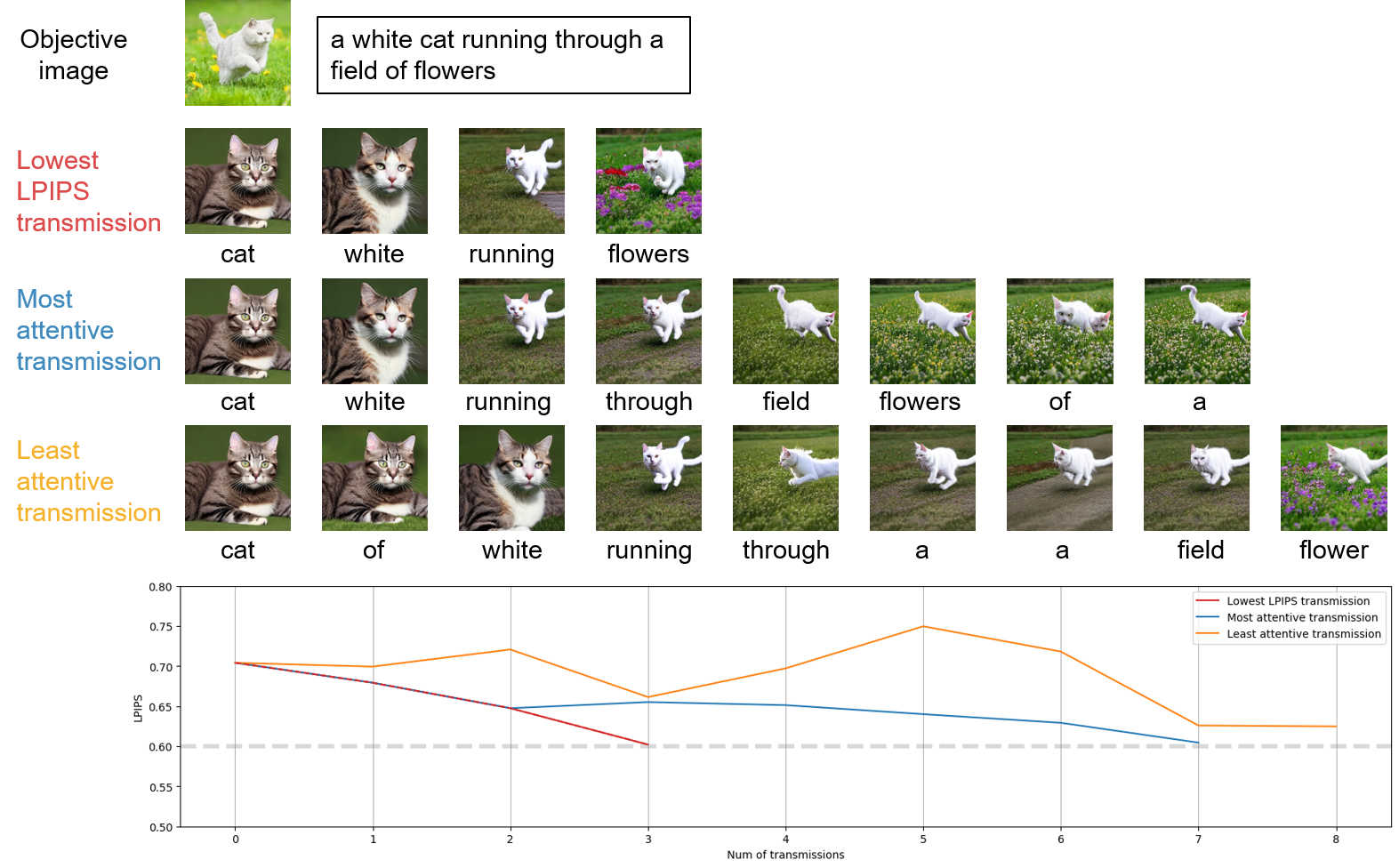}
    \caption{A simulation result comparing three types of sequential text transmission scheme. Selected word of Alice to transmit per communication steps, resulting generated image of Bob with the received words, and corresponding LPIPS loss.}
    \label{fig: ex}
\end{figure*}

\section{Schemes of Generative Multi-Modal models}
\subsection{Image-to-Text generation}
Known as image captioning, image-to-text generation has evolved with the advancement of language foundation models. They act as image-grounded text decoders using Transformer-based text prompt generation techniques. It processes one caption per image by conditioned with image representations produced by vision transformers (ViT\cite{dosovitskiy2020image}). We use a notable model, openAI's BLIP\cite{li2022blip} and it utilizes Bert\cite{devlin2018bert} as LLM.  

\subsection{Text-to-Image generation} \label{section: T2I}

Text-to-image model structure follows the basis of image generative model, which generates realistic images from random noise. One kind of popular image generative model is Latent Diffusion \cite{rombach2022high}. Latent Diffusion models synthesize an image from normally distributed variables by gradually denoising, conditioned on given text prompt. Commonly, they are composed of three modules, respectively responsible for text embedding, generating latent vectors, and decoding latent vectors to an image. A representative model, stable diffusion \cite{patil2022stable} uses CLIP\cite{radford2021learning}, VAE\cite{kingma2013auto} and U-Net\cite{ronneberger2015u} each. U-Net is trained to learn an image data distribution by repeatedly denoising an noised image and reverse process, adding normally distributed noise to the image through Markov chain process for $T$ steps each. In image generation process, the U-Net $\epsilon_\theta$ denoises from pure Gaussian noise $l_T ~ N(l_T;O,I)$ to a high-fidelity image for $T$ gradual steps, following joint distribution with former vectors as below.
\[p\theta(l_0;T) = p(l_T)\prod_{t=1}^{T}p_\theta(l_{t-1}|l_t)\]
\[p_\theta(l_{t-1}|l_t) = N(l_(t-q); \frac{1}{\sqrt{1-\beta_t}}(l_t + \beta_t\epsilon_\theta(l_t,t)),\beta_t I) ,\]
where $\{\beta_t\}_{t=1}^{T}$ is a variance schedule, and latent vectors is a normally distributed noise at step $T$ and are purely denoised at step $0$. Here a text prompt is conditioned to the process. CLIP text encoder $\tau_\phi$ produces a text representation $\tau_\phi(y)$ with given text $y$ and injected to cross-attention layers of U-Net. Then key-value pairs ($K,V$) are made by projecting text representation and query ($Q$) by $i$th intermediate representation of U-Net $\epsilon_i(l_t)$. Then attention value is computed with value, text, assigned to weight, calculated by compatibility with query and key, both modals \cite{vaswani2017attention}.
\[Attention(Q,K,V) = softmax(\frac{QK_\tau}{\sqrt{d} })\cdot V \,\,\, \text{, where}\]
\[Q=W_Q\cdot \epsilon_i(l_t), K=W_K\cdot \tau_\phi(y), V=W_V\cdot \tau_\phi(y)\]
Finally, decoder model of VAE converts or up-sizes latent $\l_0$ to an image.


\subsection{Learned Perceptual Image Patch Similarity (LPIPS)}
In image generative models, LPIPS is well used as similarity measure between reference image and synthesized image. It quantifies closeness in perceptual aspect by using a standard pre-trained model. Two images process through specific layers of VGG\cite{simonyan2014very} and calculates a scaled and channel-wise averaged euclidean distance between two representations. The output loss value ranges from $0$ to $1$, each indicates exactly the same and not close in perceptual meanings respectively. 

\section{Semantic Sequential communication via text-to-image transformation}

In this paper, we propose a communication framework that conveys text semantics. A transmitter converts the objective image into a text, and sends it. Then a receiver reconstructs the image, based on each multi-modal generative models. This significantly reduces the communication load rather than transmitting raw image, even keeps high-fidelity between objective image and its reconstructed image.

To further enhance communication efficiency, we propose \textit{Semantic sequential communication}. The transmitter transmits the text in several divided steps, with corresponding word containing partially divided information. Each words in a sentence has own syntactic role, so when all words are gathered, they finally complete an entire meaning contained in the sentence. 

To reduce the similarity distance for at least communication steps, we fully utilize the multi-modal generative models to figure out importance or roles of each word in the sentence during the generation process. We suggest several methods to sort words in the order of essential information from the receiver's perspective. In other words, the transmitter sequentially send words in an order that prioritizes expected similarity between the objective image and an image the receiver will reconstruct with the sent words.

\section{Sequential transmission in Semantic importance-Based Priority}

\subsection{Sharing common knowledge}
To best transmit the word that Bob needs to generate as much similar as possible in early communication steps, Alice has better know the information of Bob's text-to-image process. Therefore, along with the assumption that both Alice and Bob caches and stores models they need, Alice can also store the Bob's model (Fig. \ref{fig: systemmodel}). By doing so, Alice can in advance predict the image Bob will generate based on the transmitted words. 

Depending on the environments of its device, transmitter Alice might only be affordable to store certain amount of additional model parameters other than its necessary image-to-text model. There are two probable circumstances; Alice can cache all of Bob's model, or part of it. With each circumstances, the method to select sequence of word to send differs. 

\subsection{Transmitter caches entire model of Receiver}
Alice fully utilizes the whole ability of the receiver Bob. Alice predicts the resulting generated image of the receiver Bob with text generated by her. In advance to transmitting, Alice first examine Bob's expected results with all words in the generated text prompt. For communication-efficient sequential communication, Alice selects one word that will result in the best similarity with its objective image, which we refer to as \textit{`lowest LPIPS transmission'} method. LPIPS is our regularizer for evaluating the similarity. Alice keeps transmitting words one at a time after examination. This can guarantee the communication efficiency with reaching high performance earlier.

\subsection{Transmitter caches part of model of Receiver} \label{section: scheme}
Text-to-image generative models commonly process three steps described in Section \ref{section: T2I}, text embedding with language model (LM)\cite{radford2021learning}, image generation with U-Net\cite{ronneberger2015u}, and post-processing with VAE\cite{kingma2013auto}. Among three models of Bob, Alice can examine word importance by only caching LM. The attention modules of LM feature how the model first interpret the text, specifically the relevance between words in the text. Attention value is multiplication of attention weights ($softmax(\frac{QK}{\sqrt{d}})$) and value ($V$) described in Section \ref{section: T2I}. The attention weights are shaped in a square matrix in each number of attention layers and number of heads per one layer. The square matrix has dimensions of $(text length) \times (text length)$, so each pixel shows how closely two words of x-axis and y-axis are related. Alice guesses the most attentive word among the entire text for the first communication step, and after, she guesses relative word between the previously sent word.

We suggest two possible methods to sequentially select words, choosing the most and the least attentive words with the last sent word. We compare two methods, namely \textit{most attentive transmission} and \textit{least attentive transmission}, along with `lowest LPIPS transmission'. The more attentive words can give richer information related to last word, and commonly main objects or verbs explaining the image are evaluated as attentive to most of words in the text. The least attentive words also are expected to give different information than the former words, providing nearly complete information of the image in early communication steps.

\section{Simulation Results}
\subsection{Experiment settings}
Alice, the transmitter, used BLIP\cite{li2022blip}, while Bob, the receiver, used Stable Diffusion v2\cite{patil2022stable}. Here we assume LPIPS under 0.60 as a successful communication. The transmitter transmits each word at a time step until the accumulated sequence of transmitted words produces a similar image and reaches the target LPIPS.

\subsection{Sequential text transmission scheme}
We compared three methods in how fast they reach the target performance. The goal is to send at least number of words to get lower than LPIPS $0.6$ to spend at least number of communication steps. Figure \ref{fig: ex} demonstrates the selected word of Alice to transmit per communication steps, and resulting generated image of Bob with the received words, and corresponding LPIPS loss. Bob achieves lower LPIPS loss than the target in faster steps in order of \textit{`lowest LPIPS transmission'}, \textit{`most attentive transmission'} and \textit{`least attentive transmission'} methods by Alice.

\subsection{Lowest LPIPS transmission}
In case of the most superior `lowest LPIPS transmission', the transmitter first simulate with all candidate words with Bob's model to pick one that results in the most lowest LPIPS loss. Although it costs lots of computation load for simulation, Alice can be aware of the exact performance in advance, and transmit word with the best performance (lowest LPIPS loss). Therefore it can reach the target most quickly.

\subsection{Most attentive transmission and Least attentive transmission}
On the other hand, in \textit{most attentive transmission} and \textit{least attentive transmission} methods, Alice cannot expect the exact performance in advance of transmitting a word, because she has only a part of Bob's model unavailable to simulate with the whole model. 

The figure shows that sending the most attentive words along with the last sent word is more beneficial in communication term than sending the least attentive words. From the LPIPS results and even a human perspective on the transmitted word in each steps, it can be observed that \textit{`most attentive transmission'} predominantly selects crucial words in the early stages of communication, resulting in images that are more semantically similar. This means the attention weights of language model of Bob's text-to-image models show key nouns or verbs are representing the image, and relevant words also help describing the rich image. In contrast, the \textit{`least attentive transmission'} tends to favor less impactful conjunctive words such as `a' or `of', which have weaker roles in the image. Moreover, since it consecutively selects weakly associated words, the combination constructs nonsense meaning, leading to an increase in LPIPS loss.


\section{Concluding Remarks}
This paper proposes and compares sequential text transmission schemes for the text-to-image generation task. We divide the scenarios into two cases: when Alice has entire Bob model parameters and part of it. For each case, we suggest how the text should be composed and transmitted. We demonstrate that more efficient communication can be achieved when Alice predicts the result, even though it needs high memory and computation to operate with entire Bob's model. However, even when Alice only has a part of the model, it is still possible to achieve good performance using a language attention module.


\section*{Acknowledgment}
This work was supported by Institute of Information \& communications Technology Planning \& Evaluation (IITP) grant funded by the Korea government(MSIT) (No.2021-0-00347, 6G Post-MAC (POsitioning- \& Spectrum-aware intelligenT MAC for Computing \& Communication Convergence)), and (No.2021-0-00270, Development of 5G MEC framework to improve food factory productivity, automate and optimize flexible packaging)


\begin{thebibliography}{10}
\providecommand{\url}[1]{#1}
\csname url@samestyle\endcsname
\providecommand{\newblock}{\relax}
\providecommand{\bibinfo}[2]{#2}
\providecommand{\BIBentrySTDinterwordspacing}{\spaceskip=0pt\relax}
\providecommand{\BIBentryALTinterwordstretchfactor}{4}
\providecommand{\BIBentryALTinterwordspacing}{\spaceskip=\fontdimen2\font plus
\BIBentryALTinterwordstretchfactor\fontdimen3\font minus
  \fontdimen4\font\relax}
\providecommand{\BIBforeignlanguage}[2]{{%
\expandafter\ifx\csname l@#1\endcsname\relax
\typeout{** WARNING: IEEEtran.bst: No hyphenation pattern has been}%
\typeout{** loaded for the language `#1'. Using the pattern for}%
\typeout{** the default language instead.}%
\else
\language=\csname l@#1\endcsname
\fi
#2}}
\providecommand{\BIBdecl}{\relax}
\BIBdecl

\bibitem{ramesh2021zero}
A.~Ramesh, M.~Pavlov, G.~Goh, S.~Gray, C.~Voss, A.~Radford, M.~Chen, and
  I.~Sutskever, ``Zero-shot text-to-image generation,'' in \emph{International
  Conference on Machine Learning}.\hskip 1em plus 0.5em minus 0.4em\relax PMLR,
  2021, pp. 8821--8831.

\bibitem{radford2021learning}
A.~Radford, J.~W. Kim, C.~Hallacy, A.~Ramesh, G.~Goh, S.~Agarwal, G.~Sastry,
  A.~Askell, P.~Mishkin, J.~Clark \emph{et~al.}, ``Learning transferable visual
  models from natural language supervision,'' in \emph{International conference
  on machine learning}.\hskip 1em plus 0.5em minus 0.4em\relax PMLR, 2021, pp.
  8748--8763.

\bibitem{li2022blip}
J.~Li, D.~Li, C.~Xiong, and S.~Hoi, ``Blip: Bootstrapping language-image
  pre-training for unified vision-language understanding and generation,'' in
  \emph{International Conference on Machine Learning}.\hskip 1em plus 0.5em
  minus 0.4em\relax PMLR, 2022, pp. 12\,888--12\,900.

\bibitem{dosovitskiy2020image}
A.~Dosovitskiy, L.~Beyer, A.~Kolesnikov, D.~Weissenborn, X.~Zhai,
  T.~Unterthiner, M.~Dehghani, M.~Minderer, G.~Heigold, S.~Gelly \emph{et~al.},
  ``An image is worth 16x16 words: Transformers for image recognition at
  scale,'' \emph{arXiv preprint arXiv:2010.11929}, 2020.

\bibitem{devlin2018bert}
J.~Devlin, M.-W. Chang, K.~Lee, and K.~Toutanova, ``Bert: Pre-training of deep
  bidirectional transformers for language understanding,'' \emph{arXiv preprint
  arXiv:1810.04805}, 2018.

\bibitem{rombach2022high}
R.~Rombach, A.~Blattmann, D.~Lorenz, P.~Esser, and B.~Ommer, ``High-resolution
  image synthesis with latent diffusion models,'' in \emph{Proceedings of the
  IEEE/CVF Conference on Computer Vision and Pattern Recognition}, 2022, pp.
  10\,684--10\,695.

\bibitem{patil2022stable}
S.~Patil, P.~Cuenca, N.~Lambert, and P.~von Platen, ``Stable diffusion with
  diffusers,'' \emph{Hugging Face Blog}, 2022,
  [https://huggingface.co/blog/rlhf](https://huggingface.co/blog/stable\_diffusion).

\bibitem{kingma2013auto}
D.~P. Kingma and M.~Welling, ``Auto-encoding variational bayes,'' \emph{arXiv
  preprint arXiv:1312.6114}, 2013.

\bibitem{ronneberger2015u}
O.~Ronneberger, P.~Fischer, and T.~Brox, ``U-net: Convolutional networks for
  biomedical image segmentation,'' in \emph{Medical Image Computing and
  Computer-Assisted Intervention--MICCAI 2015: 18th International Conference,
  Munich, Germany, October 5-9, 2015, Proceedings, Part III 18}.\hskip 1em plus
  0.5em minus 0.4em\relax Springer, 2015, pp. 234--241.

\bibitem{vaswani2017attention}
A.~Vaswani, N.~Shazeer, N.~Parmar, J.~Uszkoreit, L.~Jones, A.~N. Gomez,
  {\L}.~Kaiser, and I.~Polosukhin, ``Attention is all you need,''
  \emph{Advances in neural information processing systems}, vol.~30, 2017.

\bibitem{simonyan2014very}
K.~Simonyan and A.~Zisserman, ``Very deep convolutional networks for
  large-scale image recognition,'' \emph{arXiv preprint arXiv:1409.1556}, 2014.

\end{thebibliography}

\vspace{12pt}

\end{document}